\input{aipcheck}

\edef\optionlist{%
   \variorefoptionifavailable        
   draft,%
   \psnfssproblemoption              
   tnotealph}

\documentclass[\optionlist]{aipproc}

\newcommand{\be}{\begin{equation}}
\newcommand{\ee}{\end{equation}}
\newcommand{\bea}{\begin{eqnarray}}
\newcommand{\eea}{\end{eqnarray}}

\def\Journal#1#2#3#4{{#1} {\bf #2}, #3 (#4)}

\def\NPB{{Nucl. Phys.} B}

\def\PLB{{Phys. Lett.} B}
\def\PRL{Phys. Rev. Lett.}
\def\PRD{{ Phys. Rev.} D}

\layoutstyle{6x9}

\usepackage{ttct0001}
\usepackage{graphicx}

\usepackage{shortvrb}
\MakeShortVerb\|

\makeatother

\begin{document}

\begin{flushright}
LMU 01/11\\
August 2001
\end{flushright}

\author{Alexander Khodjamirian \footnote{
\it on leave from Yerevan Physics Institute, 375036 Yerevan, Armenia}}{
address={Sektion Physik der Universit\"at M\"unchen,
Thereseinstr. 37, D-80333 M\"unchen, Germany},
  email={Alexander.Khodjamirian@cern.ch}}

\title{QCD Sum Rules for Heavy Flavour Physics}

\date{}

\begin{abstract}
Uses of QCD sum rules for heavy flavoured hadrons
are discussed. "Standard" applications such as the 
determination of the $b$, $c$ quark masses, the calculation of $f_B$, $f_D$
and of the heavy-to-light form factors are overviewed.
Furthermore, a new approach to calculate the $B\to \pi\pi$
hadronic matrix elements from QCD light-cone sum rules is described.
\end{abstract}

\maketitle

\begin{center}
presented at the International Workshop on QCD: Theory and
Experiment,\\ Martina Franca (Italy), June 2001
\end{center}

\section{Introduction}

Starting from the first work \cite{SVZ} the method of QCD sum rules
was frequently applied to various problems of heavy flavour physics.
Nowadays, different versions of the sum rule approach are used, all of them
based on the general idea of calculating a quark-current correlation
function and relating it to the hadronic parameters
via dispersion relations.

The original version \cite{SVZ} (often called
{\em SVZ sum rules}) employs the operator-product expansion (OPE) of
correlation functions in terms of quark and gluon vacuum condensates.
A typical and important application of this technique
is the calculation \cite{6auth,RRY,fB} of the $B$-meson decay
constant $f_B$ defined as
$\langle 0\mid m_b\bar{q}i\gamma_5 b\mid B\rangle=f_Bm_B^2$, $q=u,d$.
One starts from the correlation function of two heavy-light
currents
\be
\Pi(q^2)=
i \int d^4xe^{iqx}\langle 0\mid T\{m_b\bar{q}i\gamma_5 b(x),
m_b\bar{b}i\gamma_5 q(0)\}\mid 0\rangle\,.
\label{fBcorr}
\ee
Depending on the region of the momentum transfer $q$, the amplitude
$\Pi(q^2)$ 
represents either a short-distance fluctuation (at $q^2$ far below $m_b^2$) 
or a complicated sum over hadronic states (at $q^2 \geq m_B^2$)
starting from the ground-state $B$ meson.
At  $\mid q^2 -m_b^2\mid \gg \Lambda_{QCD}^2$,
the correlation function (\ref{fBcorr}) is
approximately calculated in terms
of the condensate expansion including the perturbative part
(the loop and $O(\alpha_s)$ correction) and the quark-, gluon- and
quark-gluon condensate contributions. A detailed derivation and
the resulting expression can be found, e.g.  in Ref.~\cite{KR}. 
On the other hand, the correlation function (\ref{fBcorr})
obeys
the dispersion relation, schematically
\be
\Pi(q^2)=
\frac{f_B^2m_B^4}{m_B^2-q^2}
+\sum\limits_{B_h}
\frac{f_{B_h}^2m_{B_h}^4}{m_{B_h}^2-q^2}\,,
\label{dispfB}
\ee
where the contribution of the ground-state $B$ meson is shown
explicitly and the sum over $B_h$ represents
the excited resonances and the continuum of hadronic states
with the $B$ meson quantum numbers.
Matching the condensate expansion of $\Pi(q^2)$
with the dispersion relation one obtains the primary sum rule.
To achieve practical results one may proceed in different directions.
Knowing the values of universal QCD parameters such as
$\alpha_s$, $m_b$ and the condensate densities $\langle G^2\rangle$,
$\langle \bar{q}q \rangle$ 
it is possible to estimate 
$f_B$ applying the quark-hadron duality approximation
to the contribution of higher states in Eq.~(\ref{dispfB}).
A reverse  way to use QCD sum rules is available in cases
when the parameters of few lowest hadronic
states in the dispersion relation are measured.
Saturating the hadronic part of the sum rule it is then possible
to determine the QCD parameters. This kind of analysis is accessible
for the two-point correlation functions
of $\bar{b}\gamma_\mu b$ or $\bar{c}\gamma_\mu c$
currents, where the hadronic states are $\Upsilon$- or
$\psi$-resonances, respectively,  with measured decay constants and masses.

A different version of QCD sum rules,
the so called {\em light-cone sum rules} (LCSR) \cite{lcsr} are
attracting a lot of  attention in recent years.  LCSR are
used to calculate various hadronic transition matrix elements. The outline
of this method can be illustrated taking 
the  $B\to \pi $ form factor calculation \cite{Bpi} as an example.
The correlation function  is in this case a vacuum-to-pion matrix element
\bea
F_\lambda (p,q)=
i\!\! \int \!d^4xe^{ipx}\langle \pi^+(q)\!\mid T\{\bar{u}\gamma_\lambda b(x),
m_b\bar{b}i\gamma_5 d(0)\}\mid\! 0\rangle
\nonumber
\\
= F((p+q)^2,p^2)q_\mu+
\tilde{F}((p+q)^2,p^2)p_\mu\,,
\label{lcsrcorr}
\eea
correlating the $b \to u$ weak current and the heavy-light current
interpolating $B_d$ meson.
At large virtualities, that is at $\mid (p+q)^2-m_b^2 \mid \gg \Lambda_{QCD}^2$
and $p^2\ll m_b^2$ this correlator can be calculated employing
OPE near the light-cone $x^2=0$. The result is expressed in terms of pion distribution
amplitudes (DA) of growing twists, the most important one being the
twist-2 pion DA defined as \cite{exclusive}
\be
\langle \pi^+(q)\mid\bar{u}(x)\gamma_\mu\gamma_5 d(0) \mid 0 \rangle
=-iq_\mu f_\pi \int\limits^1_0 du \,e^{iuqx}~\varphi_\pi(u)\\.
\label{phipi}
\ee
Higher-twist contributions are suppressed by inverse
powers of $ ((p+q)^2-m_b^2) $.
Again, using the dispersion relation in the channel of the $B$-meson
current, one obtains
\be
F((p+q)^2,p^2)=
\frac{2f_B f^+_{B\pi}(p^2)m_B^2}{m_B^2-(p+q)^2}
+\sum\limits_{B_h}\frac{2f_{B_h} f^+_{B_h\pi}(p^2)m_{B_h}^2}{m_{B_h}^2-(p+q)^2}~,
\label{displcsr}
\ee
where the ground-state contribution contains a product
of $f_B$ and the $B\to \pi$ form factor $f^+_{B\pi}(p^2)$ defined
in the standard way: 
$
\langle \pi^+(q)\mid \bar u \gamma_\mu b \mid \bar{B}^0(p+q)\rangle =
2 f^+_{B\pi}(p^2) q_\mu + ...$.
Matching the result of the twist expansion
for the amplitude $F$ with the dispersion relation
one calculates the $B\to \pi$ form factor
provided that $f_B$ is obtained from the SVZ sum rule
as explained above. Conversely, one may use LCSR
to estimate the parameters of the light-cone DA by
saturating the sum-rule relations with the experimentally known form factors
and decay constants. This is possible, e.g. for
various vacuum-to-pion correlators of light-quark currents yielding
LCSR for $\gamma \gamma^* \to \pi  $ \cite{AK}
or for the pion e.m. form factor \cite{BKM}.
Again, quark-hadron duality is employed in 
the same way as in SVZ sum rules.  
Note that in LCSR the meson DA (such as
$\varphi_\pi(u)$ with its normalization factor $f_\pi$)
play the role of nonperturbative inputs,
similar to the role the vacuum condensates play in SVZ sum rules.

For completeness, one has to mention that an independent
version of QCD sum rules emerges when one takes the local limit of LCSR.
The result is equivalent to the
{\em sum rules in the external field} \cite{ext}. 
In particular, for the pion-to vacuum correlation function (\ref{lcsrcorr})
the local limit ($q \to 0$) corresponds to the external
soft-pion field.

One should always bear in mind that QCD sum rules  have a limited
accuracy. Both OPE and the duality approximation should be kept under
control by performing the Borel transformation or
taking the power moments. The working region
of the corresponding auxiliary parameters (Borel parameter or the
number of moment) has to be restricted, so that the contributions of excited
states and higher orders in OPE are simultaneously small.

Why QCD sum rules are in particular advantageous for  heavy
flavour physics? First of all, the presence of 
an intrinsic heavy-quark mass scale 
provides necessary conditions for applying the short-distance or light-cone
OPE to the correlation functions. One has to emphasize that no
infinite quark mass limit is necessary. The sum rules can be derived
in full QCD for finite $b$ and $c$ quark masses.
Moreover, since the correlation functions are  Lorentz-covariant objects
one does not need  nonrelativistic approximations.
On the other hand, the sum rule method is very flexible
and can be applied in the frameworks of the effective QCD theories
such as HQET and NRQCD. Finally, since QCD sum rules employ
universal inputs (condensates, light-cone DA)
there is a possibility to estimate theoretical uncertainties
in the determination of heavy-flavour parameters such as $f_B$ or $f^+_{B\pi}$.

In what follows I will overview the status
of a few "standard" applications  of QCD sum rules:
the $m_b$, $m_c$ determination, the calculation of $f_B$, $f_D$
and of the $B\to \pi,K,...$, $D\to \pi,K,...$ form factors.
Furthermore, I describe a  new application of the method, the calculation of
the $B\to \pi\pi$ matrix elements.
I will not discuss many other interesting
applications such as $B_c$ meson, $B-\bar{B}$ mixing parameter,
properties of heavy baryons, sum rules in HQET.
These issues as well as many other
details concerning QCD sum rules can be found
in the recent review \cite{CK}.

\section{"Standard" applications of QCD sum rules}

\subsection{ $b$ and  $c$ quark masses}
The heavy quark mass $m_Q$, $Q=b$ or $c$, can be determined if one considers
the two-point correlation function of two $\bar{Q}\gamma_\mu Q$ currents
and uses for the hadronic dispersion relation
the experimental data on the masses and electronic widths
of $J^{P}=1^-$ quarkonium levels, $\Upsilon$ or $\psi$ resonances,
respectively. In recent years there has been a lot of progress
in the determination of the $b$ quark mass employing precise
data on  six $\Upsilon(nS)$ resonances obtained in
$e^+e^-$-annihilation. The emphasize was made on working
with the highest possible power moments of the correlation function
in which the region of small quark velocities $v_Q=\sqrt{1-4m_Q^2/q^2}$
dominates and the NRQCD approximation is valid. In this framework it is
possible to sum over Coulomb $O(\alpha_s^n/v_Q^n$) terms and to include
relativistic corrections order by order  
(for a recent review, see e.g. Ref.~\cite{Beneke99}). 
An alternative approach 
is to stay within full QCD and to employ 
few first moments of the SVZ sum rules, determining $m_b$ in a purely
relativistic way. In this case the Coulomb resummation
is not accessible but also not that important. The price to pay
is the sensitivity
to the nonresonant tail of the hadronic spectral function in the dispersion
relation which, in principle, can
be reliably estimated from experimental data on
the inclusive $e^+e^- \to b\bar{b}$ cross section above resonances. 
In Table~1 the earlier SVZ and more recent NRQCD sum rule results are
compared with the $m_b$ determinations using other approaches. The
values of $\overline{m}_b(\overline{m}_b)$
obtained by various methods agree within uncertainties.
The potential of SVZ sum rules for $b$-quarkonium
in full QCD is not yet thoroughly
exploited, e.g. there is a possibility to include the
already available $O(\alpha_s^2)$
corrections to the correlation function \cite{chetyrkin97b}.
\begin{table}
\begin{tabular}{rrrr}
\hline
 $m_b$ (GeV) & $\overline{m}_b(\overline{m}_b)$ (GeV)& Ref.&Method\\
\hline
$4.72 \pm 0.05$ & &\cite{DomingPaver92} &  SVZ\\
$4.62 \pm 0.02$ & & \cite{Narison94} & `` \\
\hline \hline
$4.827\pm 0.007$& & \cite{Voloshin95} &   \\ 
$4.84 \pm 0.08$ &$4.19 \pm 0.06 $&\cite{JaminPich99} & \\ 
&$4.20 \pm 0.10 $&\cite{MelnikovYelkhov99} & NRQCD \\
$4.88 \pm 0.10$ &$4.20 \pm0.06 $&\cite{Hoang99} & SR \\
$4.80 \pm 0.06$ &$4.21\pm 0.11$ &\cite{PeninPivovarov98} & \\
$4.97\pm 0.17$&$4.26 \pm 0.10 $&\cite{BenekeSinger99} &  \\
\hline \hline
$5.04 \pm 0.09 $&$4.44\pm 0.04$ & \cite{PinedaYndurain98}& \\
-- &$4.24 \pm 0.09 $& \cite{BenekeSinger99} &$\Upsilon$  \\ 
--&$4.21 \pm 0.07 $& \cite{Hoang99} &+NRQCD \\ 
&$4.21^{\pm 0.09}_{\mp 0.025}$&\cite{Pineda01} &  \\
\hline \hline
--&$4.26\pm0.11$&\cite{Hashimoto00}& lattice av. \\
\hline
\caption{$b$-quark mass determination: QCD sum rules vs other methods;
$m_b$($\overline{m}_b$) is the pole ($\overline{MS}$) mass}.
\label{mb}
\vspace{-1.0cm}
\end{tabular}
\end{table}

\vspace{0.2cm}
\begin{table}[h]
\begin{tabular}{rrrr}
\hline 
$m_c$ (GeV) & ${\overline m_c}(\overline m_c)$ (GeV) &Ref.&method\\
\hline
$1.46 \pm 0.05 $ &  &\cite{DominguezGluckmanPaver94}
&SVZ        \\
$1.42 \pm 0.03 $ &$1.23^{+0.02}_{-0.04}\pm 0.03$&\cite{Narison94} & ''\\
\hline \hline 
$1.70 \pm 0.13$ &$1.23\pm 0.09$ &\cite{EidemullerJamin00}  & SVZ+NRQCD \\
\hline \hline 
-&$1.37 \pm 0.09$ & \cite{PenSchilcher01} & FESR \\
\hline \hline
- &$1.21\pm 0.07 ^{\pm 0.065}_{\mp 0.045}$ &\cite{Pineda01} & $m_B-m_D$ + HQET \\
\hline \hline
-&$1.525\pm0.040\pm0.125$&\cite{APE'98} & lattice QCD\\
&$1.33\pm0.08$&\cite{FNAL98}  & ``\\
&$1.20\pm0.04\pm0.11\pm0.2$ & \cite{NRQCD99} & latt. NRQCD\\
\hline
\caption{$m_c$ determination: QCD sum rules vs other methods.}
\label{mc}
\vspace{-0.3cm}
\end{tabular}
\end{table}
The predictions for the $c$-quark mass obtained from QCD sum rules
for the charmonium system are collected in Table~2, in comparison
with the results of various other methods. 
The most recent sum rule analysis \cite{EidemullerJamin00}
combines NRQCD at small $v_c$ with the 
full QCD spectral function at large $v_c$. The latter
includes $O(\alpha_s)^2$ terms and $d=6,8$ gluon condensates.
Although the value of $\overline{m}_c(\overline{m}_c)$ is in
agreement with earlier sum rule determinations,  the pole $c$ quark mass is
surprisingly large, being interpreted \cite{EidemullerJamin00}
as a result of large Coulomb corrections. At this point one has
to note that for the charmonium levels the applicability
of NRQCD is questionable due to the large average values of $v_c$.
An update of the full QCD SVZ sum rule at the $O(\alpha_s^2)$ level
remains a task which deserves attention.
Summarizing the estimates given in Tables 1,2 I will adopt
the following intervals of the pole quark masses obtained from QCD sum rules:
\be
m_b=4.8 \pm 0.1~ \mbox{GeV}, ~~m_c= 1.3 \pm 0.1 ~\mbox{GeV}.
\label{masses}
\ee

\subsection{$f_{B}$ and $f_D$} 

Having determined the value of $m_b$ with a certain accuracy it is possible to
calculate $f_{B}$ from the SVZ sum rule based
of the dispersion relation (\ref{dispfB}).
For the $D$-meson decay constant $f_D$ the analogous sum rule 
is obtained by a simple $b\to c$ ($\bar{B}\to D$) replacement in the $f_B$ sum rule,
together with the necessary adjustment of the normalization scale.
Moreover, switching to $q=s$ in Eq.~(\ref{fBcorr}) it is possible to
predict also the ratios $f_{B_s}/f_{B}$ and $f_{D_s}/f_{D}$. 
Currently, the OPE of the correlation function (\ref{fBcorr})
includes the $O(\alpha_s)$ correction to the perturbative part which is
rather large and all  $d\leq 6$ condensate contributions.

The values of $f_{B}$ and $f_D$ determined from SVZ sum rules
are quite sensitive to the $b$ and $c$ quark pole masses.
Varying the latter in the intervals (\ref{masses}) one
typically obtains (see, e.g. the review \cite{CK}):
\bea
f_B= 170 \mp 30~\mbox{MeV},~~   f_D=180 \mp 30~\mbox{MeV}\,, 
\label{fBD}
\nonumber
\\
f_{Bs}/f_B=1.16\pm 0.09 \,, ~~f_{D_s}/f_D=1.19\pm 0.08\,.  
\eea
Here, the normalization scales   
$\mu^2_b\sim m_B^2-m_b^2$ and  $\mu^2_c\sim m_D^2-m_c^2$, respectively
are adopted. These scales are of the order of the corresponding
Borel parameters reflecting the average virtualities of the quarks
in the correlators.
Within uncertainties, the predictions (\ref{fBD})
agree  with the lattice determinations of the heavy meson decay constants.

Are improvements in $f_{B,D}$ determination
still possible? Apart from narrowing the intervals for
heavy quark masses one has to cope with the numerically
large $O(\alpha_s)$ correction. Studies of $O(\alpha_s^2)$ effects,
at least for an accurate fixing of the relevant scale in the
$O(\alpha_s)$ term are necessary.
An important progress in this direction 
is the calculation of the three-loop radiative
corrections to the heavy-to-light correlator
completed just recently \cite{Chetyrkin2001}.
The accuracy of the sum rule could be improved further if
the $d=7$ corrections proportional to the heavy quark
mass are calculated including
both factorizable $\sim\langle G^2\rangle\langle \bar{\psi}\psi\rangle $ and
nonfactorizable $d=7$ condensates. On the hadronic side, it
is important to get a better control over quark-hadron
duality in the $B$ and $D$ channels.
For the latter channel a valuable information could be provided
by experimental studies of excited $D$ states
in the semileptonic $B\to X_c l \nu$ decays. Knowing the positions of
excited $D$ resonances one may try various
alternative patterns for the hadronic spectral function
in the $f_D$ sum rule and
thereby test the validity and consistency of the
duality approximation.

\subsection{Heavy-to-light form factors}

The procedure of obtaining LCSR for the $B\to \pi$ form factor is
briefly outlined
in the Introduction. More detailed discussion can be found
in Ref.~\cite{KR}. The most recent LCSR prediction \cite{KRWWY00}
for $f_{B\pi}^+$ is presented in Fig.~1.
This calculation includes twist 2 (LO and $O(\alpha_s)$ NLO)
and twist 3,4 effects. The inputs used to calculate
$f^+_{B\pi}$ from LCSR are: 1)  the values of $f_B,m_b$ and the duality threshold, all taken from the SVZ sum rule for $f_B$ and
2) the pion DA of twist 2,3,4.
The shapes of the latter are largely fixed by their asymptotic forms
whereas the sensitivity of LCSR to the nonasymptotic effects in DA
turns out to be very mild.

\begin{figure}
\includegraphics[height=0.38\textwidth,angle=0]{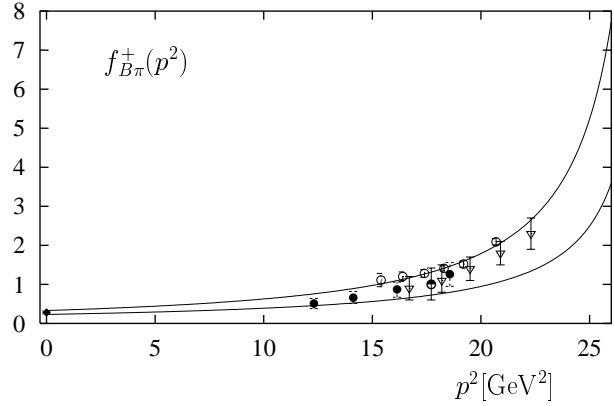}
\caption{
{\it  The LCSR prediction for the $B\to \pi$ form factor $f^+$
\cite{KRWWY00}. The full curves 
indicate the theoretical uncertainty, the points
represent various lattice QCD calculations. }}
\end{figure}
The LCSR result \cite{KRWWY00} is parametrized in a
form suggested in Ref.~\cite{BK}:
\be
f_{B\pi}^+(p^2) =\frac{f_{B\pi}^+(0)}{(1-p^2/m_{B^*}^2)(1-\alpha_{B\pi}p^2/m_{B^*}^2)}
\label{Bpi}
\ee
with $ f^+_{B \to \pi}(0)=0.28 \pm 0.05$ and
$ \alpha_{B\pi}=0.32^{+0.21}_{-0.07}$.
The theoretical uncertainties are estimated by varying
all sum rule parameters within allowed regions and adding them up
linearly.
The accuracy of the form factor calculation can still be improved
if the twist 3 $O(\alpha_s)$ correction is calculated and if
the pion DA are better constrained, e.g. from LCSR for the pion form factors
or from lattice QCD studies.

With  the form factor (\ref{Bpi})
it is possible to calculate the semileptonic
$\bar{B}^0\to \pi^+ l \bar{\nu}_l$ width ($l=e,\mu$) and to
extract the CKM parameter $\mid V_{ub}\mid$
using the current experimental data:
$BR(B\to \pi l \nu)= (1.8 \pm 0.4 \pm 0.4)\cdot 10^{-4}$ (CLEO \cite{CLEO}) and
$BR(B\to \pi l \nu)= (1.28 \pm 0.20 \pm 0.26)\cdot 10^{-4}$
(Belle, preliminary \cite{Belle}). The results are:\\ 
$
\mid V_{ub}\mid  =(4.0 \pm 0.6 \pm 0.7)\cdot 10^{-3} \mbox{(CLEO)}
$, and 
$
\mid V_{ub}\mid  =(3.4 \pm 0.4 \pm 0.6)\cdot 10^{-3} \mbox{(Belle)}
$,\\
where the first error is  experimental  and the second one is
caused by the theoretical uncertainty of the form factor calculation.

Replacing the pion with the kaon in the correlation function (\ref{lcsrcorr})
leads to LCSR for the
$B\to K$ form factor, including the effects of the $SU(3)$-flavour
symmetry breaking such as $f_K/f_\pi \neq 1$ and the  asymmetry in the kaon
DA $\varphi_K(u)$. Interestingly, the predicted ratio
\cite{KRWWY00}
\be
f^+_{B \to K}(0)/f^+_{B \to \pi}=1.28^{+0.18}_{-0.10} \,,
\ee
is mainly sensitive to the value of the strange quark mass
$m_s( 1 \mbox{GeV}) = 150 \mp 50$ MeV. This result indicates
that the rate of $SU(3)$ breaking
could be quite noticeable, an important message for studies
of CP-violation in hadronic $B$ decays where various SU(3) relations
are frequently used.
The semileptonic form factors in the charmed sector are also
predicted from LCSR , e.g. \cite{KRWWY00} : $ f^+_{D \to \pi}(0)=0.65 \pm 0.11$
and $ f^+_{D \to K}(0)/f^+_{D \to \pi}=1.20$ (at
$m_s (1~\mbox{GeV})=150$ MeV), 
in a good agreement with both experiment and lattice QCD.

To complete our discussion on heavy-to-light form factors one has to mention 
various $B\to V$ form factors, $V=K^*,\rho,\phi$, relevant for
$B\to V l \nu_l $ and $B\to V \gamma$ decays. Their most  accurate
calculation is in Ref.~\cite{BallBraun}.
Using LCSR it is also possible to estimate the
amplitudes of $B\to \rho \gamma$ weak annihilation \cite{KSW,AB} 
and the $B\to \mu \nu \gamma$ width \cite{KSW} employing the photon DA. 
The list of heavy-to-light semileptonic and radiative processes treated
with the help of LCSR can be enlarged
to include also $B\to a_{0,1,2}$, $B\to K_{0}^*,K_1,K_2^*$
transition form factors if the corresponding DA 
of these light mesons are worked out. Another potentially
interesting application is to employ the two-pion DA \cite{twopion}
in studying $B\to \pi\pi l \nu$ decay.
The first step in this direction was done in Ref.~\cite{Maul}.

\subsection{ $D^* D\pi$-coupling, QCD sum rules vs experiment}

Recently the total width of $D^*$ meson was measured by CLEO
collaboration \cite{CLEODstar}: 
$\Gamma_{tot}(D^*)= 96 \pm 4 \pm 22$ keV. This remarkable measurement
yields the strong $D^*D\pi$ coupling 
$g_{D^*D\pi} = 17.9 \pm 0.3 \pm 1.9$ defined as in Ref.~\cite{BBKR}.
Among many theoretical predictions for this coupling obtained by
various methods I would like to single out the LCSR prediction \cite{BBKR}
updated in Ref.~\cite{KRWYcoupl} by including the $O(\alpha_s)$ correction
to the twist 2 term: $g_{D^*D\pi} = 10 \pm 3.5 $.
The sum rule is derived \cite{BBKR} from the correlator (\ref{lcsrcorr})
employing the double dispersion relation. An estimate in
the same ballpark is 
obtained from the QCD sum rules in the soft pion limit \cite{BBKR,Colang}.
Note that the theoretical
uncertainty quoted above includes variation of all inputs
within reasonable limits,
therefore it is difficult to push the LCSR prediction above its
upper limit $g_{D^*D\pi} = 13.5 $
which is still  25\% lower than the central value of the CLEO
measurement.
If the currently observed discrepancy between the LCSR prediction and
experiment persists in future one might suspect 
that the simple quark-hadron duality ansatz which works in the
one-variable dispersion relations is too crude for the double
dispersion relation. 

Let me make one parenthetical remark.
It is often claimed that knowing 
the value of the $D^*D\pi$ coupling one fixes the effective
scale-independent coupling in HQET defined as
$\hat{g}=f_\pi g_{H^*H\pi}/2m_H$, $H=B,D$.
However, in the charmed sector there are 
large $1/m_H$  corrections to the HQET limit.
Indeed, as shown in Refs.~\cite{BBKR,CF98}
where both $D^*D\pi$ and $B^*B\pi$ couplings
are calculated from LCSR, they can be fitted
to a single effective $\hat{g}$ only by adding a substantial
$1/m_H$ correction:  $g_{H^*H\pi}= 2m_H\hat{g}/f_\pi(1+\Delta/m_H)$,
with $\Delta \sim 1~\mbox{GeV}$. Therefore, expressing
$g_{D^*D\pi}$ in terms of
$\hat{g}$ is not a straightforward procedure.

\section{Light-Cone Sum Rules for Hadronic $B$ Decays}

The CP-violation studies are nowadays
concentrated
on the two-body hadronic $B$ decays. In order to fully
explore experimental data  one needs reliable theoretical
predictions on  hadronic matrix elements
of the operators $O_i$ entering  the effective weak Hamiltonian,
$H_W= \frac{G_F}{\sqrt{2}}\sum _i \lambda_i^{CKM}c_i(\mu) O_i$. The
solution of this tremendously difficult problem can only
be achieved within approximate QCD methods, such as the
recently developed QCD factorization \cite{BBNS}.

Here I will shortly outline a new approach to this problem \cite{AK2001}
which is based on LCSR and allows
to calculate the hadronic matrix elements 
in the same framework as the $B\to \pi$ form factor.
As a study case the matrix elements
$\langle \pi^-\pi^+ \mid O_{1,2} \mid B \rangle_{Emission}$
of the current-current
operators $O_{1,2}$  for $\bar{B}^0\to \pi^+\pi^-$
in the emission topology
are considered, where $O_1=(\bar{d}\Gamma_\mu u)(\bar{u}\Gamma^\mu b)$
and $O_2$ is replaced by a combination of  $O_1$ and the colour-octet
operator
$
\widetilde{O}_1=(\bar{d}\Gamma_\mu \frac{\lambda^a}2u)(\bar{u}\Gamma^\mu
\frac{\lambda^a}2 b)
$.

As usual  in the sum rule derivation, one constructs a 
suitable correlation function:
\be
F^{(O)}_\alpha (p,q,k)=\! -\!\!\int\!\! d^4x e^{-i(p-q)x} \!\!\int\!\! d^4y e^{i(p-k)y}
\langle 0 \mid T\{\bar{u}\gamma_\alpha\gamma_5 d(y)O(0)
m_b\bar{b}i\gamma_5d(x)\}\mid \pi^-\!(q)\rangle\,.
\label{Bpipicorr}
\ee
Here the effective operator $O=O_1$ or $\widetilde{O}_1$
is correlated with the currents
interpolating $B$ meson and pion. In the above, a fictive momentum $k$ is
attributed to the weak vertex to avoid certain technical difficulties in 
the dispersion relations. Furthermore, we put $p^2=k^2=0$ and consider
the kinematical region of large spacelike external momenta 
$\mid (p-k)^2\mid \sim \mid (p-q)^2 \mid \sim \mid P^2 \mid
\gg \Lambda^2_{QCD}$, where $P= p-k-q$.
Due to the large $b$ quark mass it is possible to apply
the light-cone OPE to the correlator (\ref{Bpipicorr})
in this region.
The lowest-order diagram for the operator $O_1$ is shown
in Fig. 2a. It factorizes into a simple light-quark loop and
the vacuum-to-pion correlation function similar to Eq.~(\ref{lcsrcorr}).
The OPE of the correlator (\ref{Bpipicorr})
with the operator $\widetilde{O}_1$ starts from the
diagrams containing a one-gluon nonfactorizable exchange: 
either a soft (low virtuality) gluon
which is absorbed by the pion DA (Fig.~2b)
or a hard gluon exchanged between
the light-quark loop and the heavy-light part of the correlator.
The latter effect corresponds to the $O(\alpha_s)$ two-loop diagrams 
which demand technically difficult calculation. In what follows
we concentrate on the soft nonfactorizable effect represented
by the diagrams in Fig.~2b. Their calculation involves twist
3 and 4 quark-antiquark-gluon DA of the pion. The key nonperturbative
parameters determining these DA are the matrix elements
$
\langle 0\mid \bar{u}
\sigma_{\mu\nu}\gamma_5g_s G_{\alpha\beta}d \mid \pi\rangle$ and
$
\langle 0 \mid g_s\bar{u}\tilde{G}_{\alpha\beta}\gamma^\mu d \mid \pi \rangle
$
estimated from SVZ sum rules \cite{parameters}.

\begin{figure}
\includegraphics[height=0.35\textwidth,angle=0]{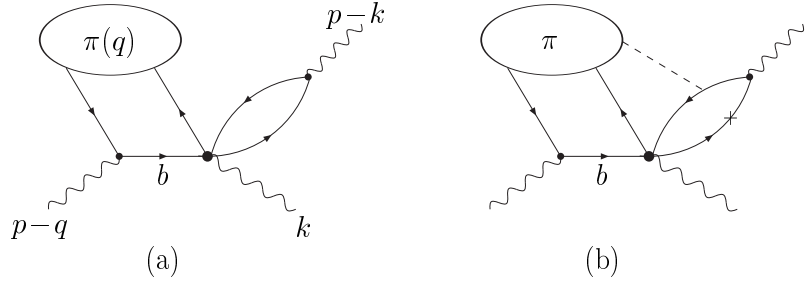}
\caption{ {\it Diagrams corresponding to the leading
order of the correlator (\ref{Bpipicorr}) for $O=O_1,
\widetilde{O}_1$; the cross indicates the point of gluon emission
in the second similar diagram.}}
\end{figure}
The procedure of the sum rule derivation from Eq.~(\ref{Bpipicorr})
is more complicated
than in the $B\to \pi$ form factor case. It can be
shortly summarized as follows:

1. The dispersion relation in the pion-current channel with the
momentum $(p-k)$ is employed together with the quark-hadron duality
approximation, allowing one to obtain an analytic expression for the
hadronic matrix element
\be
\Pi^{(O)}_{\pi\pi}((p-q)^2,P^2)=i\int d^4xe^{-i(p-q)x}
\langle \pi^-(p-k)\mid T\{O(0)m_b\bar{b}i\gamma_5d \}\mid \pi^-(q)\rangle\,. 
\label{matr}
\ee
This matrix element resembles the pion form factor at large
spacelike momentum transfer $P^2$ where, instead of a simple e.m. vertex, one
has a more complicated short-distance part with a virtual $b$ quark. 

2. Analytic continuation of Eq.~(\ref{matr})
in the variable $P^2$ to the large timelike $P^2= m_B^2$ is performed.
This procedure is analogous  to the transition from large spacelike
to large timelike momentum transfers for the pion e.m.
form factor. Note that an imaginary part may emerge
as a result of this continuation. It has to be interpreted
as a strong final state interaction phase.

3. The dispersion relation for $\Pi^{(O)}_{\pi\pi}((p-q)^2,m_B^2)$
in the variable $(p-q)^2$ (in the $B$-meson channel) is written down
and the duality ansatz for the higher $B$ states is applied.
As a final result, one obtains
LCSR for the on-shell $ \bar{B}_d\to\pi^+\pi^-$ matrix
element of the operator $O$ where the fictive momentum $k$
vanishes due to $P^2=m_B^2$.
As usual, in order to suppress the higher states and to reduce the sensitivity
to the duality approximation in both pion and $B$ meson channels,
two independent Borel transformations
are performed in the variables $(p-k)^2$ and $(p-q)^2$, respectively.
Note that  all parameters entering the obtained
sum rules are fixed either from the SVZ sum rules 
for two-point correlation functions or
from LCSR for $f^+_{B\pi}$.  

The resulting sum rule for the matrix element
of $O_1$ in the leading order 
simply factorizes into a product of SVZ sum rule
for $f_\pi$ and the LCSR for $B\to \pi$ form factor:
$
\langle \pi^-\pi^+ \mid O_1 \mid B \rangle_{E}
=if_\pi f_{B\pi}^+(0)_{LCSR}~m_B^2\,,
$
thereby reproducing the factorization approximation
in the limit of the heavy quark mass. 
The LCSR for the matrix element of the colour-octet operator $\widetilde{O}_1$
calculated from the diagrams in Fig.~2b quantifies the soft
nonfactorizable effect. Importantly, at $m_b\to \infty$
it is $1/m_b$ suppressed with respect to the factorizable
part, in accordance with QCD factorization \cite{BBNS}.
Numerically, 
\be
\frac{\langle \pi^-\pi^+ \mid\widetilde{O}_1 \mid B \rangle}
{\langle \pi^-\pi^+ \mid O_1 \mid B \rangle}\equiv \frac{\lambda_E}{m_B}
,~~\lambda_E= 50 \div 150~ \mbox{ MeV},
\ee
that is, the soft nonfactorizable effect due to $\widetilde{O}_1$
is small and does not contain imaginary part. At the same time,
the soft effect is as important as the small $O(\alpha_s)$
hard nonfactorizable effects calculated in the QCD factorization approach.

The $\bar{B}_d \to \pi^+\pi^-$ decay amplitude obtained from LCSR  
\bea
{\cal A}(\bar{B}_d\to \pi^+\pi^-) =i\frac{G_F}{\sqrt{2}}
V_{ub}V^*_{ud}f_\pi [f_{B\pi}^+(0)]_{LCSR}~m_B^2 
\Bigg\{c_1(\mu) +\frac{c_2(\mu)}{3}
\nonumber
\\
+2c_2(\mu) \Big(\frac{\lambda_{E}}{m_B}+ O(\alpha_s)\Big)\Bigg\}+ ...
\eea
has to be completed by calculating
the $O(\alpha_s)$ nonfactorizable  effects. After including the latter
in the decay amplitude the scale $\mu$ dependence has to be partially
canceled and the imaginary  part will emerge at $O(\alpha_s)$.
It is also important to calculate one by one
the contributions of annihilation, penguin topologies
for $O_{1,2}$  as well as the matrix elements
of penguin operators $O_n$, $n\geq 3$ denoted by ellipses in the above.
The LCSR approach can be generalized
to  other channels such as  $B\to K\pi,KK, D\pi, J/\psi K$.

\section{Summary } 
The aim of this minireview was to demonstrate that
employing QCD sum rules one is able to determine various 
heavy-flavour parameters starting from the most
fundamental ones, the heavy quark masses,
and ending with the most complicated ones, the
hadronic matrix elements of nonleptonic $B$ decays. The method
is selfsufficient, that is, extracting a certain parameter
from a sum rule, one uses the result 
in the other sum rules to calculate more complicated parameters. The following
hierarchy can be traced:
$$
m_b \to f_B\to f^+_{B\pi} \to \langle \pi\pi \mid O_i\mid B\rangle~. 
$$
Summarizing, QCD sum rules remain a reliable approximate approach 
well equipped to attack various topical problems in the physics
of heavy flavours.\\

\noindent {\bf Note added:}\\
After this meeting, two new QCD sum rule calculations of
the heavy meson decay constants
have been published, both including the $O(\alpha_s^2)$ correction
\cite{Chetyrkin2001} to the heavy-light correlator. The first analysis is done
in HQET\cite{PeninStein} and predicts $f_B=206 \pm 20$ MeV,
$f_D=195 \pm 20$ MeV . The second one \cite{JaminfB} uses $\overline{MS}$-mass
of the $b$ quark instead of the pole mass yielding
$f_B=197 \pm 23$ MeV and $f_{B_s}=232 \pm 25$ MeV.
Within uncertainties, both results \cite{PeninStein,JaminfB}
agree with the intervals in Eq.~(\ref{fBD}).\\ 

\noindent {\bf Acknowledgements}. I am grateful to P. Colangelo and G. Nardulli
for an opportunity to participate in this fruitful and enjoyable workshop.\\

\end{document}